# Age-specific contacts and travel patterns in the spatial spread of 2009 H1N1 influenza pandemic.

Andrea Apolloni[1*], Chiara Poletto[2,3,4*], Vittoria Colizza[3,4,5†]

1) Department of Infectious Disease Epidemiology, London School of Hygiene and Tropical Medicine, London, United Kingdom
2) Computational Epidemiology Laboratory, Institute for Scientific Interchange (ISI), Torino, Italy
3) INSERM, U707, Paris, France
4) UPMC Université Paris 06, Faculté de Médecine Pierre et Marie Curie, UMR S 707, Paris, France
5) Institute for Scientific Interchange (ISI), Torino, Italy

*These authors contributed equally to the work
†corresponding author: vittoria.colizza@inserm.fr

## Abstract

**Background.** Confirmed H1N1 cases during late spring and summer 2009 in various countries showed a substantial age shift between importations and local transmission cases, with adults mainly responsible for seeding unaffected regions and children most frequently driving community outbreaks.

**Methods.** We introduce a multi-host stochastic metapopulation model with two age classes to analytically investigate the role of a heterogeneously mixing population and its associated non-homogeneous travel behaviors on the risk of a major epidemic. We inform the model with demographic data, contact data and travel statistics of Europe and Mexico, and calibrate it to the 2009 H1N1 pandemic early outbreak. We allow for variations of the model parameters to explore the conditions of invasion under different scenarios.

**Results.** We derive the expression for the potential of global invasion of the epidemic that depends on the transmissibility of the pathogen, the transportation network and mobility features, the demographic profile and the mixing pattern. Higher assortativity in the contact pattern greatly increases the probability of spatial containment of the epidemic, this effect being contrasted by an increase in the social activity of adults vs. children. Heterogeneous features of the mobility network characterizing its topology and traffic flows strongly favor the invasion of the pathogen at the spatial level, as also a larger fraction of children traveling. Variations in the demographic profile and mixing habits across countries lead to heterogeneous outbreak situations. Model results are compatible with the H1N1 spatial transmission dynamics observed.

**Conclusions**. This work illustrates the importance of considering age-dependent mixing profiles and mobility features coupled together to study the conditions for the spatial invasion of an emerging influenza pandemic. Its results allow the immediate assessment of the risk of a major epidemic for a specific scenario upon availability of data, and the evaluation of the potential effectiveness of public health interventions targeting specific age groups, their interactions and mobility behaviors. The approach provides a general modeling framework that can be used for other types of partitions of the host population and applied to different settings.

# Background

The data collected during and after the 2009 H1N1 pandemic has contributed to achieve major insights regarding key factors of the transmission dynamics of the novel strain of influenza. Two aspects emerging from surveillance and serological data during the initial phase of the outbreak became strikingly clear: (i) international movements of passengers by air travel drove the spatial dissemination of the pathogen at the global level [1-3], and (ii) initial local epidemics mainly occurred in schools [2,4-7]. Country surveillance data from summer 2009 [8-13] show a dramatic difference in the age distributions of imported cases and indigenous cases, with imported cases on average older than the ones generated by local transmission (see Figure 1A). This result strengthens the previous observations and highlights the presence of an age shift between the seeding events established by travelers from affected areas – mainly adults – and the local outbreaks mainly occurring among school-aged children. Travel statistics indeed confirm that the vast majority of passengers flying are aged $\geq 18$ years (see Figure 1B).

The role of children or adults in driving a major epidemic can be assessed with a simplified modeling approach expressed into two age classes and quantifying the probability of temporary extinction of seed individuals in each class depending on their mixing patterns [14]. Here we aim at fully addressing the interplay between the two factors emerged as empirical evidence during the initial outbreak of H1N1 pandemic – namely, age-specific seeding events and age-specific epidemic dynamics – by considering a spatially explicit model that integrates the mobility patterns of flights connecting different populated areas, as well as age-dependent travel behavior in addition to age-specific mixing. We introduce a multi-host stochastic metapopulation model that includes heterogeneous features in: the spatial structure of the population; the travel behavior of individuals depending on their age; the corresponding mixing patterns. We assume a simple two age-groups classification as this allows the analytical treatment of the model to obtain the expression of the conditions for a major epidemic in terms of the age-specific contacts and travel features. By exploring theoretical contact matrices with varying assortativity levels (i.e. within-group mixing), we assess the role of assortativity on the risk of a major epidemic and compare it to scenarios informed with estimates from the H1N1 pandemic. We find that the assortativity observed in real data tend to drive the system to extinction, and is counterbalanced by the heterogeneity of the air mobility network structure, by the travel behavior of adults and by the relative proportion of contacts established by adults, all aspects that favor the virus spread. Through a systematic exploration of the role of the various ingredients considered in the system, the presented results allow the risk assessment analysis of a specific epidemic scenario and can be extended to other infectious diseases where the population partition may play a relevant role.

## Methods

**Demographic and travel data**

We consider a metapopulation framework to simulate the spread of an infectious disease across subpopulations of individuals through mobility connections. The approach is generically applicable to various real-world systems and here we focus on modeling an emerging influenza pandemic across urban areas through air travel. We consider the distinct cases of 8 countries in Europe, and Mexico, for which data needed to inform the model are available.

The network specifying the coupling between different populations in real systems is in many cases very heterogeneous, and examples range from transportation infrastructures to mobility patterns of various type [15-19]. In the case of air travel the coupling is given by the direct flights connecting different airports and the number of passengers flying on those connections. Analyses of air transportation data have shown the presence of large variability in the number of connections per airport, and of broad fluctuations in the traffic handled by each airport or flowing on a given connection between origin and destination [15,16]. In the Additional File we report the example of the European air transportation network [20], showing the probability distributions of the number of connections $k$ per airport (i.e. the subpopulation's degree) and of the flows of passengers $w_{ij}$ travelling between any pair of linked airports $i$ and $j$. These flows can also be expressed in terms of the number of connections of the origin/destination airports, where the average weight $\langle w_{ij} \rangle$ along the link connecting airports $i$ and $j$ is a function of their degrees $k_i$ and $k_j$, $\langle w_{ij} \rangle \propto (k_i k_j)^\theta$, with $\theta \cong 0.5$ in the worldwide air-transportation network [15]. Such features, obtained from empirical evidence, will be used in the following to define the metapopulation model, by creating a realistic synthetic network of mobility connecting a number $V$ of urban areas.

Since we are interested in exploring how non-homogeneous travel habits, coupled with non-homogeneous mixing patterns, may drive the conditions for the spatial invasion of an epidemic, we collected age travel statistics across different countries. These are typically obtained from travel surveys at airports, and collect a variety of information about passengers and their travel behavior including demographic data. Figure 1B shows the percentage of travelers in the younger age class for a variety of sources and for different airports in various countries. If we consider a classification of the younger age group that includes individuals up to 18-21 years old (where the upper value of the range depends on the specific limits imposed by the age classification adopted by each survey), the fraction of children traveling by air is on average equal to 7%, with a maximum variation between 0.7% (observed for Teheran airport, Iran) and 9.2% (Luton airport, UK). Other statistical sources have larger age brackets, with no breakdown below 25 or 30 years old, as shown in the Figure. In these cases, we can still estimate the fraction of traveling individuals in the age class [0-18] years old, by rescaling the statistics assuming a constant ratio across countries between the percentages of travelers in the [0-18] years old and those of travelers in the [0-24] or [0-29] years old classes, as obtained from sources with a finer age classification. The estimated values are reported in the caption of Figure 1B and are consistent with the data.

Demographic data for the age distribution of the population by country was obtained from Eurostat [20] for European countries, and from the U.N. database [21] for Mexico. The data is provided by yearly age groups for European countries and 5 years age groups for Mexico, and it allows the calculation of the fraction of people under a given age, corresponding to the classification used in the model. If we consider the younger age class up to 18 years old, we obtain relative sizes of the population ranging in Europe from 17.1% for Italy to 22.2% for United Kingdom and Luxembourg, with an average value of 19.7%. The corresponding value in Mexico increases to 32.3% [21], with a classification up to 15 years old for data availability reasons.

**Theoretical and data-driven age mixing patterns**

In addition to the travel behavior and demographic features of the population, we need to consider the mixing pattern among population classes. For the purpose of the study, we consider data-driven mixing patterns by country [2,14,22] and we also define a theoretical contact matrix, dependent on a set of parameters that we vary in a range of plausible values in order to systematically explore the behavior of the global invasion under different mixing conditions.

We consider the population divided into two classes, children and adults identified by subscripts $c$ and $a$, respectively. Our aim is indeed to characterize the impact on the invasion dynamics of the two coupled phenomena, namely the early phase of the epidemic outbreak locally driven mainly by children and the spatial dissemination of the epidemic mainly driven by the adults traveling. While demographic and travel data allow for the consideration of a larger number of age classes, our choice of two classes is motivated by the simpler parameterization of the model in terms of the mixing patterns that allows us to formulate the invasion conditions analytically.

Children are assumed to represent a fraction $\alpha$ of the population $N$, so that the population size of children is expressed as $N_c = \alpha N$; the population size of adults being thus expressed as $N_a = (1-\alpha)N$. $\alpha$ is a parameter defined between 0 and 1, which we assume to be homogeneous across the metapopulation system. This is a plausible assumption if we consider a metapopulation model applied to a single country or to regions that are rather homogeneous in terms of their demography, as these values are not expected to substantially vary. In the following sections we will explore different values of $\alpha$, and compare these results to the ones obtained in the case the model is informed with the data-driven values of $\alpha$ from the demographic statistics of the European countries under study and of Mexico.

We define the contact matrix $C = \{C_{ij}\}$ capturing the mixing between different age classes as

$$\begin{pmatrix} C_{cc} & C_{ca} \\ C_{ac} & C_{aa} \end{pmatrix} = \begin{pmatrix} p_c q_c \frac{N}{N_c} & (1-p_a)q_a \frac{N}{N_c} \\ (1-p_c)q_c \frac{N}{N_a} & p_a q_a \frac{N}{N_a} \end{pmatrix}, \qquad (1)$$

where $q_c$ and $q_a$ are the average number of contacts per unit time established by individuals in the children and adult classes, respectively, and $p_c$ and $p_a$ are the fractions of contacts that occur between individuals of the same age class. The variables $q_c$ and $q_a$ represent a measure of social activity of the individuals, whereas $p_c$ and $p_a$ describe how these contacts are established among classes. A variety of different assumptions can be done to describe the social mixing pattern, as the variables of the contact matrix of the above expression are not uniquely defined and need to be parameterized through available demographic and serologic data [23]. Here we focus our attention on assortative mixing, indicating the tendency of individuals in a given class to preferably interact with other individuals of the same class. This is indeed observed in real data where contacts patterns are found to be highly assortative with age [22,24,25], with a remarkable similarity across different European countries [22]. By indicating with $\varepsilon_c$ ($\varepsilon_a$) the average fraction of contacts that a child (adult) establishes with an adult (child), i.e. across age groups, we can simply express $p_c$ ($p_a$) in terms of $\varepsilon_c$ ($\varepsilon_a$) as $p_c = 1 - \varepsilon_c$ ($p_a = 1 - \varepsilon_a$), thus the contact matrix can be rewritten as

$$\begin{pmatrix} C_{cc} & C_{ca} \\ C_{ac} & C_{aa} \end{pmatrix} = q_c \begin{pmatrix} \frac{1-\varepsilon_c}{\alpha} & \frac{\eta \varepsilon_a}{\alpha} \\ \frac{\varepsilon_c}{1-\alpha} & \frac{\eta(1-\varepsilon_a)}{1-\alpha} \end{pmatrix}, \quad (2)$$

where we have used the relation $N_c = \alpha N$ and have indicated with $\eta$ the ratio between the average contact numbers per age class, $\eta = q_a/q_c$. Interactions are reciprocal such that the number of contacts between children and adults is the same as the number of contacts between adults and children, requiring the matrix to be symmetric, i.e. $C_{ca} = C_{ac}$ or $\varepsilon_c \alpha = \eta \varepsilon_a (1-\alpha)$. We indicate with $\varepsilon = \varepsilon_c \alpha = \eta \varepsilon_a (1-\alpha)$ the total fraction of contacts across age classes, so that the contact matrix finally reads

$$\begin{pmatrix} C_{cc} & C_{ca} \\ C_{ac} & C_{aa} \end{pmatrix} = q_c \begin{pmatrix} \frac{\alpha-\varepsilon}{\alpha^2} & \frac{\varepsilon}{\alpha(1-\alpha)} \\ \frac{\varepsilon}{\alpha(1-\alpha)} & \frac{\eta(1-\alpha)-\varepsilon}{(1-\alpha)^2} \end{pmatrix}. \quad (3)$$

The matrix is fully expressed in terms of the average number of contacts in the children class, $q_c$, the fraction of children $\alpha$, the ratio $\eta$ between the average number of contacts in the adult class and the one in the children class, and the parameter $\varepsilon$. The latter is defined between 0 and $\min\{\alpha, \eta(1-\alpha)\}$, and regulates the degree of assortativity between the age classes [26], with $\varepsilon \to 0$ indicating high assortativity and $\varepsilon \to \min\{\alpha, \eta(1-\alpha)\}$ indicating low assortativity, as schematically shown in Figure 2. Table 1 reports the list of variables and parameters used to introduce the age classes. A more general modeling framework that include also other mixing pattern types and that is applicable to different social classifications besides age is the object of future work.

To compare the theoretical results to realistic situations, we estimate the parameters $\varepsilon$ and $\eta$ from data-driven contact matrices $C = \{C_{ij}\}$, obtained from a population-based prospective survey of mixing patterns in eight European countries using a common paper-diary methodology [22]. We used smoothed daily contact rates based on a bivariate smoother [27] defined for age classes of 1 year interval, relative to all contacts, i.e. including both physical and non-physical contacts [22] (more details can be found in the Additional File). The European countries considered include: Belgium (BE), Germany (DE), Finland (FI), Great Britain (GB), Italy (IT), Luxembourg (LU), The Netherlands (NL), and Poland (PL).

In addition we also informed our model with the mixing patterns for Mexico, obtained from studies on the early outbreak of the 2009 H1N1 pandemic in the country [2]. The values for the parameters $\{\alpha, \eta, \varepsilon\}$ obtained for all countries under study are reported in Table 2.

**Spatial metapopulation model with age structure**

We consider a population of individuals that is spatially structured into $V$ subpopulations coupled by human mobility patterns, representing a metapopulation network where nodes correspond to subpopulations where the infection dynamics takes place, and links correspond to the mobility processes among them. We initially present this approach considering only one class of individuals, to highlight its main features and the assumptions we make, and in the following we will introduce the age structure of the population. The model is informed with the statistical laws empirically observed in real data on human population and mobility by air-travel, and discussed in the Demographic and travel data subsection. The metapopulation structure is characterized by a random connectivity pattern described by an arbitrary degree distribution $P(k)$. In the following we will explore the role of realistic heterogeneous network structures, adopting power-law degree distributions $P(k) \propto k^{-\gamma}$ for analytical convenience, mimicking in this way the airline network as drawn from realistic data. Following the scaling properties observed in real-world mobility data, we define the number of individuals moving from the subpopulation of degree $k$ to the subpopulation of degree $k'$ as $w_{kk'} = w_0 (kk')^\theta$. We fix the exponent $\theta$ to 0.5 and the scaling factor $w_0$ to 1.0, based on the empirical findings [15]. Smaller values of $w_0$ are also explored to simulate the implementation of travel-related intervention strategies such as reductions of the travel flows following the start of the outbreak.

We model the travel diffusion process to match the patterns $w_{kk'}$, assuming that travelers are randomly chosen in the population with the per capita diffusion rate $d_{kk'} = w_{kk'}/N_k$ where $N_k$ is a variable indicating the population size [28].

The variables introduced to define the metapopulation model solely depend on the degree $k$ of each subpopulation, therefore we introduce a degree-block notation [29] that assumes statistical equivalence for subpopulations of equal degree [30]. It corresponds to assuming that all subpopulations having the same number of connections are considered statistically identical regarding the features of the metapopulation system relevant for the mobility and disease spreading processes (such as for instance the population size and the traffic of passengers). While disregarding more specific properties of each individual subpopulation – that may be related for instance to local, geographical or cultural aspects – this mean field approximation is able to account for the large degree fluctuations empirically observed, to capture the degree dependence of the system's properties as found in the data, and also to allow for an analytical treatment of the system's behavior [30]. The full list of variables used to define the metapopulation model with one class of individuals is provided in Table 3.

The model defined so far considers a single class of individuals who homogeneously mix in the population. Here we introduce the age structure in order to consider different mixing patterns and travel probabilities depending on the age class, based on the results presented in the previous subsections. We consider a simple SIR compartmental model to describe the infection dynamics of the influenza epidemic, where individuals are assigned to mutually exclusive compartments – susceptible (S), infectious (I), and recovered (R) individuals [23]. Susceptible individuals may contract the infection from infectious individuals and enter the infectious compartment; all infectious individuals then recover permanently and enter the recovered compartment. The disease dynamics is encoded in the next generation matrix $\boldsymbol{R} = \{R_{ij}\}$ i.e. the average number of secondary infections in age group $j$ generated by a single primary case in age group $i$ in a fully susceptible population, with $i,j \in \{c,a\}$. We assume that the infectious period is exponentially distributed (with average value $\mu^{-1}$) and that both transmission rate and recovery rate are independent of the age group, thus neglecting age-specific susceptibility or infectiousness, as in [14]. We consider a fully susceptible population, and also the case of an age-specific prior immunity, as detailed in the next subsection. If we assume that disease transmission may only occur along the contacts captured by the matrix $\boldsymbol{C} = \{C_{ij}\}$ [24,31], we can express the generic entry of the next generation matrix as $R_{ia} = \frac{\beta}{\mu}C_{ia}\alpha$ and $R_{ic} = \frac{\beta}{\mu}C_{ic}(1-\alpha)$ [32,33] with $i \in \{a,c\}$ and $C_{ij}$ the contact matrix defined in Eq.(1). In the application of the contact matrix of Eq.(3) to the metapopulation model, we assume that all parameters used to define the partition of the population into age classes are independent of spatial features, being therefore constant across subpopulations (e.g. the children population fraction of a subpopulation with degree $k$ is given by $N_{k,c} = \alpha N_k$). Similarly, we assume that the age-specific travel behavior does not change with the subpopulation of the system. The travel behavior is thus modeled by rescaling the per-capita diffusion rates $d_{kk'}$ with the corresponding age-specific probability of travel and the population sizes of the age classes. If we indicate with $r$ the fraction of children traveling (see Table 1), the per-capita diffusion rate for children traveling from a subpopulation with degree $k$ to a subpopulation with degree $k'$, $d_{kk',c}$, is expressed as

$$d_{kk',c} = r\frac{w_0(kk')^\theta}{N_{k,c}} = rd_{kk'}\frac{N_k}{N_{k,c}} = \frac{rd_{kk'}}{\alpha}. \tag{4}$$

Analogously, the per-capita diffusion rate for adults is given by $d_{kk',a} = (1-r)d_{kk'}/(1-\alpha)$. This allows us to consider different traveling rates depending on the age classes, and to explore a range of values of $r$, including $r = 0$, i.e. only adults travel, in addition to its real values obtained from travel statistics.

In the Additional File we also consider a more refined compartmentalization that incorporates a latency period of duration $\tau^{-1}$ to account for the time elapsing from exposure to infectiousness. This corresponds to a more accurate approximation for the description of the disease etiology of influenza. Moreover, it also allows us to extend the applicability of our modeling framework to other infectious diseases where this period may last several days, therefore being non-negligible, as in the case of the severe acute respiratory syndrome (SARS) [34].

**2009 H1N1 pandemic case study**

To clarify the impact of the study's findings in a practical situation, we apply our framework to the case study of the 2009 H1N1 pandemic influenza. We parameterize the metapopulation model to the available epidemiological estimates of the outbreak, to the demographic and travel statistics, and to the contact pattern data. We assume a fully susceptible population, and consider values of the reproductive number $R_0$ consistent with the estimations obtained for the pandemic through several methods. In particular, we focus on the range from $R_0 = 1.05$, corresponding to the lower bound of the estimate obtained from global modeling approaches for the countries in the Northern hemisphere during summer, once seasonal rescaling is taken into account [3], to $R_0 = 1.2$ corresponding to the estimates available for Japan [35] and obtained from genetic studies [2], up to $R_0 \in [1.4, 1.6]$ as estimated from the early outbreak data of the H1N1 pandemic [2].

The consideration of $R_0 = 1.05$ is also important for two additional reasons. First, this value is used here to provide a comparison between the effects that school holidays may have had in the transmission scenario in Europe during Summer 2009 due to the altered contact pattern with respect to school term [36,37]. Indeed, contact data in the UK collected during school term and school holiday periods for 2009 showed that changing mixing patterns resulted in a decrease of approximately 25% to 35% in the reproductive number of influenza during the holidays, considering physical or conversational contacts, respectively [36]. With an estimated $R_0$ for the UK in the range [1.4,1.6] when schools were open, such reductions would thus correspond to approximately $R_0 = 1.05$ during school holidays [36]. A second reason for considering the value $R_0 = 1.05$, is also for testing scenarios where a higher value of $R_0$ may have been reduced following the application of intervention strategies that do not alter the interaction or travel behaviors of individuals, such as e.g. through vaccination or antiviral treatment.

For each value of $R_0$ considered, the transmission rate $\beta$ is calculated from the dominant eigenvalue of the next generation matrix [38], where we set the infectious period $\mu^{-1}$ to 2.5 days [3].

In addition to the value $w_0 = 1$, corresponding to the estimate for the mobility scale regulating the travel fluxes obtained from the air mobility network, we also explore $w_0 = 0.5$ to simulate the travel-related controls imposed by some countries associated with the self-imposed travel limitations that contributed to a decline of about half the international air traffic to/from Mexico following the international alert in April 2009 (see [39] and references therein).

Results are obtained for eight European countries for which demographic, travel, and contact data are available, and for Mexico, this latter case informed with the age-dependent transmission matrix of Refs. [2,14]. The children age class is defined up to 18 years old for Europe and up to 15 years old for Mexico, to match available data.

We also consider the case of pre-existing immunity in older population and parameterize our model based on the serological evidence indicating that about 30 to 37% of the individuals aged $\geq 60$ years had an initial degree of immunity prior to exposure [40-42]. We assume that 33% of individuals aged $\geq 60$ years are immune and completely protected against H1N1 pandemic virus. We use the data from the different national age profiles [20] to estimate the corresponding fraction of the adult age class of each country with pre-exposure immunity.

## Results and Discussion
**Calculation of the global invasion threshold**

The reproductive number $R_0$ provides a threshold condition for a local outbreak in the community; if $R_0 > 1$ the epidemic will occur and will affect a finite fraction of the local population, otherwise the disease will die out [23]. The condition for global invasion is however made more complicated by the interplay between the local transmission dynamics and the mobility process that is responsible to seed non-infected subpopulations. Even in the occurrence of a local outbreak, given $R_0 > 1$, the epidemic may indeed fail to spread spatially if the mobility rate is not large enough to ensure the travel of infected individuals to other subpopulations before the end of the local outbreak, or if the amount of seeding cases is not large enough to ensure the start of an outbreak in the reached subpopulation due to local extinction events. All these processes have a clear stochastic nature and they are captured by the definition of an additional predictor of the disease dynamics, $R_* > 1$, regulating the number of subpopulations that become infected from a single initially infected subpopulation [28,43-46], analogously to the reproductive number $R_0$ at the individual level. An expression for $R_*$ has been found in the case of metapopulation epidemic models with different types of mobility processes, including homogeneous, traffic-driven, and population-driven diffusion rates [28,46], commuting-like processes [47,48] and origin-destination processes with adaptive behavior [49] or heterogeneous length of stay at destination [50]. In all those cases the population is assumed to mix homogeneously and to travel according to rates that are uniform across individuals. Here we go beyond those assumptions and examine the relationship between the occurrence of a global outbreak and the age-dependent transmission dynamics coupled with the age-specific travel behavior, through the calculation of the global invasion threshold $R_*$.

Let us consider the invasion process of the epidemic spread at the metapopulation level, by using the subpopulations as our elements of the description of the system. We assume that the outbreak starts in a single initially infected subpopulation of a given degree $k$ and describe the spread from one subpopulation to the neighbor subpopulations through a branching process approximation [51]. We denote by $D_k^n$ the number of infected subpopulations of degree $k$ at generation $n$, with $D_k^0$ being the initially seeded subpopulation, $D_k^1$ the subpopulations of degree $k$ of generation 1 directly infected by $D_k^0$ through the mobility process, and so on. By iterating the seeding events, it is possible to describe the evolution of the number $D_k^n$ of infected subpopulations as follows:

$$D_k^n = \sum_{k'} D_{k'}^{n-1}(k'-1)P(k|k')\left(1 - \sum_{m=0}^{n-1}\frac{D_k^m}{V_k}\right)\cdot \Omega_{k'k}(\lambda_{k'k,c},\lambda_{k'k,a}). \tag{5}$$

The r.h.s. of equation (5) describes the contribution of the subpopulations $D_{k'}^{n-1}$ of degree $k'$ at generation $n-1$ to the infection of subpopulations of degree $k$ at generation $n$. Each of the $D_{k'}^{n-1}$ subpopulations has $k'-1$ possible connections along which the infection can spread. The infection from $D_{k'}^{n-1}$ to $D_k^n$ occurs if: (i) the connections departing from nodes with degree $k'$ point to subpopulations with degree $k$, as ensured by the conditional probability $P(k|k')$; (ii) the reached subpopulations are not yet infected, as indicated by the probability $\left(1 - \sum_{m=0}^{n-1}\frac{D_k^m}{V_k}\right)$, where $V_k$ is the number of subpopulations with degree $k$; (iii) the outbreak seeded by $\lambda_{k'k,c}$ and $\lambda_{k'k,a}$ infectious individuals, children and adults, respectively, traveling from subpopulation $k'$ to subpopulation $k$ takes place, and the probability of such event is given by $\Omega_{k'k}(\lambda_{k'k,c},\lambda_{k'k,a})$. The latter term is the one that relates the dynamics of the local infection at the individual level to the coarse-grained view that describes the disease invasion at the metapopulation level. It also provides the contribution of children and adults age classes, thus including the effects of non-homogeneous travel behaviors and mixing patterns. The numbers of infectious individuals of each type flying from a subpopulation with degree $k'$ and arriving to a subpopulation with degree $k$ during the entire duration of the outbreak are given by:

$$\lambda_{k'k,c} = d_{k'k,c}\cdot\frac{z_c N_{k',c}}{\mu} = \frac{rd_{kk'}}{\alpha}\cdot\frac{z_c \alpha N_{k'}}{\mu} = rd_{kk'}\frac{z_c N_{k'}}{\mu} \tag{6}$$

$$\lambda_{k'k,a} = d_{k'k,a}\cdot\frac{z_a N_{k',a}}{\mu} = (1-r)d_{kk'}\frac{z_a N_{k'}}{\mu},$$

i.e. the final proportion $z_i$ of the $N_{k',i}$ hosts who contract the infection and diffuse with rate $d_{k'k,i}$ during their infectious period $\mu^{-1}$, with $i=c,a$. $z_c$ and $z_a$ indicate the attack rates in the children and adult age classes, respectively, and they are given by the solution to $1-z_i = \exp(-\sum_j R_{ij}z_j)$ [52]. Figure 3A shows the behavior of the final attack rates $z_c$ and $z_a$ as a function of the reproductive number $R_0$, considering the partition of the population in children and adults observed in the 8 European countries of the Polymod dataset [22] and their mixing properties. Variations between countries, depending on the age profile and the mixing patterns, are observed. Countries are generally predicted to have significantly larger epidemic sizes in children, as shown, for example, by the case of Italy. This would correspond, on average, to a larger number of individuals in the children class that could potentially seed other subpopulations and sustain the spatial invasion. However this effect is counterbalanced by the age-specific traveling probabilities that are much lower in the children class. In the case of Belgium the final epidemic sizes in the two classes are found to be almost equal, with the size of the epidemic in the adult population being slightly larger than the one in the children population. This is the only country in the dataset under consideration that has a ratio $\eta$ larger than 1, indicating a larger average number of contacts established by adults with respect to children, likely induced by the specific survey methodology adopted [22].

If we indicate with $\pi_c$ ($\pi_a$) the probability of extinction given that a single infected individual of class $c$ ($a$) is introduced in the population, the probability $\Omega_{k'k}(\lambda_{k'k,c}, \lambda_{k'k,a})$ that $\lambda_{k'k,c}$ and $\lambda_{k'k,a}$ infectious individuals traveling from subpopulation $k'$ to subpopulation $k$ would start an outbreak can be expressed as $\Omega_{k'k}(\lambda_{k'k,c}, \lambda_{k'k,a}) = 1 - \pi_c^{\lambda_{k'k,c}} \pi_a^{\lambda_{k'k,a}}$. Here the two processes are considered as independent since we assume a multi-type branching process approximation. The probabilities of epidemic extinction given the introduction of a single case are determined as the solutions of the quadratic system dependent on the elements $R_{ij}$ of the next generation matrix [14], $\pi_i = [1 - R_{ii}(1 - \pi_i) + R_{ji}(1 - \pi_j)]^{-1}$, with $i = c, a$. If $R_0 < 1$, the only solution is $\pi_c = \pi_a = 1$, i.e. the epidemic dies off. Otherwise, the system has solutions $(\pi_c, \pi_a)$ in the range [0,1], as shown in Figure 3B. All countries (except Belgium) display a larger probability of extinction related to the introduction of a single adult case in the population, with respect to the introduction of a single infectious child. Given the mixing patterns, children are therefore more likely to start an outbreak than adults. $\pi_a$ ranges between 96.5% in the case of Italy and 99% in the case of Poland for $R_0 = 1.05$, and between 83.5% and 88% for $R_0 = 1.2$, showing that there were small chances for the H1N1 pandemic outbreak to start in the summer in those countries, in agreement with the observed unfolding [53]. In general, a large dependence of the probability of extinction on the reproductive number is observed. Analogously to the behavior discussed for the epidemic size, also for the probability of extinction Belgium represents a special case for which $\pi_c$ is slightly larger than $\pi_a$, again induced by the larger average number of contacts in the adult class, differently from all other countries for which the data is available. Finally, in the case of homogeneous mixing in a non-partitioned population, we recover the probability of extinction to be equal to $R_0^{-1}$ (dashed line in the figure).

The quantities reported in panels A,B of Figure 3 represent the ingredients to assess the risk of a global epidemic as driven by the partition of the population into age classes and the non-homogeneous mixing pattern considered, as also discussed in Ref. [14]. The additional role of the non-homogeneous travel behavior is considered explicitly by modeling the invasion process through Eq. (5). To solve the system analytically, we simplify the recursive relation of Eq. (5) in the assumption of mild epidemics (i.e. in the limit of $R_0$ close to 1), introduced or emerged in the system through a localized seeding event (so that the number of infected subpopulations can be neglected at the early stage of the spatial invasion), and considering the case of a mobility network lacking topological correlations (in this approximation the conditional probability $P(k|k')$ can be simplified, see the Additional File). By manipulating the Equation (see the Additional File for the full details of the calculation), we obtain a condition allowing the increase of the number of infected subpopulations and a global epidemic in the metapopulation system only if

$$R_* = [(1-\pi_c)rz_c + (1-\pi_a)(1-r)z_a]\frac{w_0}{\mu}\frac{\langle k^{2+2\theta}\rangle - \langle k^{1+2\theta}\rangle}{\langle k\rangle} > 1, \qquad (8)$$

thus defining the global invasion threshold of the metapopulation system. Equation (8) defines the threshold condition for the global invasion: if $R_*$ assumes values larger than 1, the epidemic starting from a given subpopulation will reach global proportion affecting a finite fraction of the subpopulations of the system; if instead $R_* < 1$, the epidemic will be contained at its source and will not spread further to other locations. The global invasion threshold is a complex function of the disease history parameters, and of the parameters describing the age-specific mixing patterns and travel behavior through $\pi_c, \pi_a, r, z_c, z_a$. Its dependence on the population spatial structure is embodied by travel fluxes and the topology of the mobility network, through $w_0$, $\theta$, and the degree moments $\langle k^a \rangle$. However it is important to note that this indicator does not depend on the number of subpopulations $V$ of the system, therefore it may be applicable to a variety of countries for which data is available, independently of their size. In the following we explore the dependence of $R_*$ on these multiple factors, examine their role in driving the pandemic extinction or invasion, and provide possible applications examples for a set of countries considering the 2009 H1N1 pandemic.

**Impact of air mobility**

The term $w_0(\langle k^{2+2\theta}\rangle - \langle k^{1+2\theta}\rangle)/\langle k\rangle$ represents the contribution of the air mobility network to the global invasion threshold, with $w_0$ and $\theta$ regulating the travel fluxes, and the various moments of $k$, $\langle k^m \rangle = \sum_k k^m P(k)$, expressing the dependence on the network structure as encoded in its degree distribution $P(k) = k^{-\gamma}$. The large degree fluctuations found in real transportation systems and mobility patterns [15-19] constitute one of the mechanisms responsible for driving $R_*$ to considerable high values, even when small values of the reproductive number are considered. Figure 3C shows the dependence of $R_*$ on the reproductive number accounting for changes in the topological heterogeneity of the mobility network ($\gamma = 2$ and $\gamma = 3$) and in the traffic heterogeneity along the air connections ($\theta = 0.5$ and $\theta = 0$, the latter being the homogeneous traffic case). The larger degree fluctuations obtained in the case $\gamma = 2$ strongly increase the ratio $(\langle k^{2+2\theta} \rangle - \langle k^{1+2\theta} \rangle)/\langle k \rangle$, leading to values of $R_*$ ranging from ~10 for $R_0 = 1.05$ up to approximately $10^3$ for $R_0 = 2$, a value of the reproductive number consistent with the estimate for the 1918-1919 pandemic [54]. It is important to note that, besides the threshold condition $R_* > 1$, the absolute value of the estimator $R_*$ provides a quantitative indication of the effective reduction that needs to be reached through public health interventions in order to bring $R_*$ below its threshold value, i.e. the difference $R_* - 1$. The $R_*$ values for $\gamma = 2$ are roughly one order of magnitude greater than the values recovered in the case $\gamma = 3$, and in addition the condition for the epidemic invasion is more sensitive to variations in $R_0$ in the case of larger heterogeneities in the air mobility patterns ($\gamma = 2$ vs. $\gamma = 3$).

For both network topologies considered, the epidemic is above the threshold value of 1, and the outbreak is predicted to spread globally in the system for all diseases considered ($R_0 \geq 1.05$), consistently with the H1N1 influenza virus invasion at the global level. The partition into classes, though lowering the epidemic sizes and increasing the probability of extinction [35,55,56], is not able to drive the system below the threshold of the global invasion for the range of values explored. In addition, the contribution of the topological heterogeneity of the mobility network $(\langle k^{2+2\theta} \rangle - \langle k^{1+2\theta} \rangle)/\langle k \rangle$ in increasing the value of $R_*$ is so large that it cannot be easily counterbalanced by reductions of the mobility scale $w_0$ corresponding to interventions through air travel restrictions. This was already observed in numerical results obtained from data-driven modeling approaches and in analytical predictions based on simple homogeneous mixing among individuals within the subpopulation of the system [28,46,50,57-60]. We will see in the following subsections how differences across countries may impact the conditions for invasion, and will quantitatively assess within this framework the role of travel reductions consistent with the traffic drop observed in the traffic to/from Mexico.

Epidemic containment is instead reached for $R_0 < 1.2$ when exploring homogeneous traffic flows (i.e. $\theta = 0$ differently from the realistic scenario $\theta = 0.5$) in the case $\gamma = 3$ (see Figure 3C), showing how the large variations observed in the traffic flows along air travel connections represent an additional element favoring the spatial spread of the pathogen. On the other hand, extensive measures aimed at radically altering the amount of passenger traveling or the distribution of traffic on the air connections can hardly be achieved in reality.

**Impact of the contacts ratio $\eta$**

The global invasion threshold is a function of the extinction probabilities and of the epidemic sizes per age class for which an explicit solution cannot be obtained in the general case. An approximate solution can be recovered for small $\varepsilon$ and in the two limit cases of the contacts ratio $\eta = q_a/q_c$: $\eta \to 0$, i.e. a regime in which almost all contacts are established by children, and $\eta \to 1$, i.e. a situation that is almost homogeneous in the distribution of the number of contacts per individual. The approximate solutions are reported in the Additional File.

Besides these two limit cases, we investigate in Figure 4 the behavior of $R_*$ as a function of the contacts ratio $\eta$, exploring the interval [0.25,1] to include the estimates from the Polymod data (range from 0.62 to 0.97, Belgium excluded as discussed before) and to satisfy the existence condition on $\varepsilon$. The value of $\alpha$ is fixed to the European average, $\alpha = 0.197$, whereas each country is represented with its corresponding assortativity parameter $\varepsilon$, ranging from $\varepsilon = 0.083$ in Italy to $\varepsilon = 0.125$ in Belgium.

The global invasion threshold is predicted to increase with $\eta$, showing how a larger number of contacts established by adults, regardless of the across-groups mixing, would favor the spatial propagation. The variations observed among countries are induced by the country-specific assortativity profiles and decrease with $\eta$. Therefore, if $R_*$ reaches its critical level for relatively small values of $\eta$, for which variations across countries are still large, we may reach a heterogeneous outbreak situation in which some countries would experience spatial transmission, while others would be able to contain the outbreak simply due to the role of country-specific age profiles and of the level of assortativity in each population. This may be for example the case with $\gamma = 3$ and $R_0 < 1.2$ (Figure 4B). Results are consistent with the numerical evidence obtained from a data-driven agent-based model in Europe [61].

Countries characterized by particularly low values of $\eta$, for cultural, behavioral, and/or social reasons, would be at a lower risk of invasion. Therefore control measures aimed at reducing the contacts ratio $\eta$ of a specific country may represent an effective policy option to consider. This could be achieved through the application of workplace interventions, including for instance working at home, reducing or avoiding work meetings, and shifting the working timing to reduce overlap at workplaces and at break hours, as well as crowding on transports [62].

Finally, we report on the effects of the inclusion of the latency period in the compartmental model. The results reported in the Additional File uncover the dependence of the global invasion threshold on the generation time of the disease, i.e. the sum of the latency and infectious periods in the compartmental approximation considered. A simple addition of the latency period to the description of the disease would therefore increase the generation time and, consequently, the global threshold parameter $R_*$, while keeping the qualitative picture unchanged (see Figure S2). Individuals would indeed have a longer time span available to travel and potentially spread the disease while carrying an infection.

**Impact of assortativity and age profile**

We now examine the dependence of the critical condition for the global invasion on the assortativity level of the population partition, by plotting in Figure 5 $R_*$ as a function of the parameters $\varepsilon$ and $\eta$, where we have set the children fraction $\alpha$ equal to the European average value, $\alpha = 0.197$. $R_*$ is an increasing function of the across-groups mixing $\varepsilon$, indicating that a decrease in the assortativity level of the mixing pattern (corresponding to an increase of $\varepsilon$) favors the spatial invasion. If a larger fraction of contacts is indeed established between adults and children, the local transmission dynamics mainly driven by children is able to spread to a larger fraction of the adult population, thus increasing the chances for the spatial dissemination of the pathogen. Here we study a situation in which $r = 0$, i.e. only adults travel, in order to isolate the effect of changes in the local transmission while neglecting the role of children in the mobility process.

The rectangle shown in the panels indicates the ranges of the country values for $\varepsilon$ and $\eta$ in Europe. A variation of $\eta$ from the smallest to the largest of the country values produces a stronger effect on $R_*$ with respect to a variation of $\varepsilon$ within the European range. As such, $\eta$ represents an important source of country heterogeneity in the epidemic outcome, as already discussed.

The two panels differ for the value of the reproductive number considered that takes into account the effective reduction in the transmission potential observed during school holidays (panel A, corresponding to $R_0 = 1.05$) with respect to school term (panel B, $R_0 = 1.4$). These values are estimated for the UK on the basis of contact data for the country in the two periods [36], as illustrated in detail in the Methods section, and here we generalize this result to all countries under study to provide a comparison between school opening and school closure in terms of the predicted risk of a major outbreak. During school holiday period (panel A), European countries are found to be close to the critical level, with a portion of the parameter space for Europe in the extinction region, thus confirming the results presented before regarding variations in the country-specific risks of major epidemics. These findings are compatible with the heterogeneous transmission of H1N1 influenza virus in summer 2009 that was not sustained across continental Europe [53] and identify the main mechanisms responsible of these effects in the interplay between demographic/mixing features and the seeding during school holidays (due to similar school calendars, with the exception of the UK). If instead we consider that the influenza pandemic arrived in the UK during school term, our predictions indicate that the risk of a major epidemic with community and spatial transmission was expected to be very high even during summer period (Figure 5B), as observed in the country. Other factors undoubtedly played a role in producing the different transmission across countries, and they include humidity conditions [53], and the timing and magnitude of the coupling of the European continent to Mexico and the United States through international travel [3], both factors being country-specific and not considered here.

Additional countries with similar age profile may be mapped onto the two-dimensional plots of Figure 5 to gather immediate analytical insight regarding the risk of a major epidemic for the pathogen under consideration, given data availability on the country-specific demographic profiles and mixing habits, providing valuable predictions on the conditions for spatio-temporal transmission and informed recommendations for effective control strategies at the start of an outbreak. As an example, we also explored the situation for the United States, considering synthetic information on individual contact networks built from activity surveys and simulations for the city of Portland [25]. By assuming the validity of this synthetic information for the whole country, we can compare the global invasion threshold in the US with the one studied in Europe from real country-specific data. Similar properties are found in the air transportation networks of the two regions, and the slight differences in the age partition ($\alpha = 0.24$ in the US vs. $\alpha = 0.20$ in Europe) do not lead to great discrepancies in the behavior of the global threshold as a function of the parameter $\eta$ (see Figure S3 of the Additional File).

If we assume that children travel according to the statistics obtained from travel data (i.e. $r$ is set to 7%), the increase in $r$ leads to an increase of $R_*$, as expected, given that a fraction of the individuals driving the local outbreaks also represent potential seeds in new locations not yet affected by the epidemic (see Figure 6A). Such a small change in $r$ may also be enough to drive the system from the extinction phase to the major epidemic phase, for certain values of the other parameters. For example, an epidemic starting in a subpopulation in Italy, where the across-groups mixing is equal to $\varepsilon = 0.08$, would reach pandemic proportion if a small fraction of children would travel by air, with respect to the case in which such fraction is neglected (see the vertical line in Figure 6A). Given the specific seeding role of children vs. adults, age-targeted entry screening of travelers at airports may be envisaged in an attempt to prevent spatial transmission. The mild and self-limiting nature of most influenza infections, in addition to the presence of asymptomatic infections, however is likely to prevent a perfect identification and notification of cases at entry ports. Given that even a small percentage of children traveling would considerably increase the risk of spatial transmission, leading only to small delays in the invasion process [63], such interventions should be evaluated with respect to their actual efficacy in specific epidemic emergencies and balanced against the resources required for their implementation.

Besides mixing patterns and travel statistics, countries also differ for their age profiles (considered in the model through the parameter $\alpha$), an easy to access statistics for all countries in the world [21]. The country seed of the 2009 H1N1 pandemic, Mexico, has for instance a larger fraction of the population in the younger age class compared to the population of Europe, $\alpha = 0.32$ vs. $\alpha = 0.20$ (where the age classification used for Mexico is up to 15 years old instead of up to 18 years old as adopted in Europe, due to data availability) and is characterized by a larger number of contacts established by children with respect to adults ($\eta = 0.32$ vs. $\eta = 0.79$). Our predictions indicate that, for the same assortativity values, there exists a range of values of the epidemic transmission potential that is compatible with epidemic containment in Mexico, whereas Europe may experience spatial propagation of the disease (see Figure 6B for $R_0 = 1.05$).

Remarkably, the increase of the fraction of children and the change in the contact ratio, with no change in the across-groups mixing pattern (i.e. same $\varepsilon$), is able to drive the system in the Mexican scenario to the extinction phase, reducing of 94% the global invasion threshold obtained for Europe if we consider a pathogen with a transmission potential compatible with the seasonal estimates of the H1N1 pandemic in Europe during summer 2009 [3]. Such results would be particularly important when considering the country where a new epidemic may emerge, as predictions obtained from previous works considering hypothetical seeding countries would be hardly applicable if demographic features are different. National preparedness plans may be informed by country-specific recommendations through the present approach, and an extensive exploration of the model's results in terms of classes of demographic and contact pattern features would help in providing more general insights on classes of seeding scenarios.

In the case of $R_0 = 1.4$, i.e. the lower bound of the estimate of the reproductive number for Mexico based on the early outbreak of the 2009 H1N1 pandemic [2], our model predict that the country would be above the critical value, thus in agreement with the spatial spread that was observed.

**Effect of pre-existing immunity and travel reductions**

If we consider pre-existing immunity in the population, calculating the fraction of individuals in the adult age class that corresponds to the estimated values from serological data available after the 2009 H1N1 pandemic [40-42], we find that immunity reduces the condition for the global invasion threshold, as we could expect since a fraction of the population is now modeled to be fully protected against the virus. Figure 7 addresses the comparison between the two cases, immunity and no-immunity, by showing the two corresponding critical curves $R_* = 1$ in the $\alpha$, $\varepsilon$ plane for Europe and Mexico (panels A and B, respectively). The effect of pre-existing immunity in reducing the probability of a global epidemic spread is shown by the increase in the value of $\varepsilon$ corresponding to the invasion condition $R_* = 1$; a larger mixing across age classes is therefore needed for the pathogen to spatially propagate in case a fraction of the older age class is immune, whereas a more assortative population is predicted to be able to contain the emerging epidemic. The effect is very small for the case of Mexico, while for Europe it is more visible given that the considered European population is, on average, older than the Mexican one, and thus a larger fraction of the adult population is assumed to be immune in Europe with respect to Mexico (9.6% in Europe vs. 4.4% in Mexico).

The points $(\alpha, \varepsilon)$ parameterized with the average European data and with Mexican data (corresponding dots in the Figure 7) both lie in the spatial invasion phase when we consider $R_0 \geq 1.2$ in Europe and $R_0 \geq 1.4$ in Mexico.

As an additional factor, we also consider the effect resulting from the application of travel reductions. We simulate the travel controls applied by some countries in addition to the self-reaction of the population avoiding travel to the affected area that was observed during the early stage of the 2009 H1N1 pandemic [39], by setting $w_0 = 0.5$, i.e. a uniform reduction along all travel connections, independently of the age of travelers. Such reductions reduce the phase space of parameters leading to global invasion, as expected, however they would not be able to lead to a containment of the disease once the model is fitted to the Mexican and European data and to the H1N1 pandemic scenario, confirming previous findings [28,39,46,57-60].

**Limitations of the study**

Our study is based on a multi-host stochastic metapopulation model that considers several simplifying assumptions that we discuss in this subsection.

The partition of the population into children and adult classes is clearly very schematic, especially if we consider that finer level classifications are available for demographic data at the global scale, and for contacts data, though in a very small set of countries that are the ones considered in this study. Similar considerations arise in the case of additional heterogeneities to be included in the model, as for instance differing patterns of transmission between households, schools, and workplace settings. A higher level of structuring of the population into classes is expected to decrease the epidemic sizes per comparable groups of classes, and to increase the probability of extinction of the epidemic, with respect to our predictions (see for instance the discussions in [35,55,56] and references therein). The inclusion of such features would prevent an analytical treatment of the model and therefore push the study towards an agent-based approach [61,64,65], for which numerical simulations would represent the only available methodology.

Another assumption concerns the exponentially distributed infectious period. More realistic descriptions of the infectious period – including constant, gamma-distributed or data-driven infectious periods – were found to alter the model results by reducing the probability of extinction. Such findings were however obtained in a single population model with homogeneous mixing [66] and in a two-population model coupled by mobility, where, on the other hand, individuals were allowed only one trip, i.e. to change only one time their subpopulation [67]. Such modeling approaches and corresponding results are not applicable to our case, and a systematic understanding of the impact of the infectious period distribution on the probability of extinction in a metapopulation model is still missing.

Our analytical approach also assumes that the importation or the emergence of an infectious disease is highly localized at the beginning of the outbreak, so that it is possible to approximate the spatial spreading process in terms of a branching process evolving in a set of subpopulations not yet affected by the disease. This is similar to the approximations used to calculate the basic reproductive number in a fully susceptible population, and it is required to treat the model analytically. While this assumption can generally be considered as a good approximation to describe the early phase of an outbreak, more complicated seeding events may occur that would require numerical approaches able to explicitly take into account the initial conditions and assess the epidemic risks.

Our model is fit to demographic statistics of a set of countries and it is informed with H1N1 epidemic estimates to provide quantitative information on the risk for the pandemic invasion in such countries. However, in all our predictions we assume the same population structure ($\alpha$), contacts and mixing profiles ($\varepsilon, \eta$), and travel behavior ($r$) across all subpopulations of the metapopulation system, as informed by the data for a given country. We consider this approximation a reasonable one for regions characterized by population features that are quite uniform across space, for instance if we consider the subpopulations within a given country; already at the European level we noticed how small variations in demographic features and mixing patterns may be responsible for diverse outcomes regarding extinction or invasion, and additional layers of heterogeneities need to be considered when variations among populations within the system are larger (e.g. regarding travel behavior per age class, as shown in Figure 1b). This could be achieved by considering subpopulation-dependent variables $\{\alpha_i, \varepsilon_i, \eta_i, r_i\}$ (with $i$ indicating the subpopulation), however preventing an analytical treatment to solve the system due to the additional complications considered, thus requiring the use of numerical approaches.

Recent work relying on large-scale transmission models has explored the ability of these approaches to predict the timing of spread of the 2009 A/H1N1 influenza pandemic around the world [68]. Differently from these numerical approaches that can describe the geotemporal propagation of the infectious disease in the population, our model does not provide any temporal information on the epidemic unfolding in that all dynamical aspects are synthetically summarized in a branching process leading to the condition for global invasion. On the other hand, given the importance of demographic profiles, mixing patterns and age-specific travel resulting from the present study, it would be important to further extend large-scale spatial transmission computational models to include such features. While computationally feasible, the main limitation nowadays is represented by the availability of mixing data and travel behavior for a large set of countries, given a global level objective. This further supports the need to have multiple modeling frameworks that can complement each other in providing important information to characterize an emerging epidemic and its associated risks and impact.

Finally, we note that changes in time of population behavior as a response to the ongoing outbreak cannot be dynamically incorporated in the model. These may refer for example to changes in the contact patterns due to self-awareness or changes at the community level due to the implementation of intervention strategies to control the epidemic [49,69-73]. On the other hand, these scenarios can be separately studied with the model, assuming each of these features to be constant in time, in order to assess the effect of such changes on the corresponding risk of a major epidemic. As possible application examples we provided model predictions for different mixing patterns related to school terms and school holidays during H1N1 pandemic, as well as uniform travel restrictions that may result from self-impositions, national guidelines or travel bans. Future studies will focus on other historical epidemics, like e.g. the case of the 2002-2003 SARS outbreak, in order to explore which mechanisms, among the ones included in this approach and related to the applied interventions or to individual's self-adaptation, hindered a fully global transmission of the disease.

## Conclusions

The 2009 H1N1 pandemic represents an example of the important role that age classes have on the local and global spread of the disease during its early stage; the local outbreaks being mainly driven by children leading to epidemics in schools, whereas the adults were mainly responsible for the international dissemination by means of air travel. We introduced and solved a multi-host stochastic metapopulation model to quantify these aspects and characterize the conditions of the population partition and heterogeneous travel behavior that lead to the pandemic global invasion.

Notwithstanding the high level of assortativity observed in contact patterns data by age, that increase the probability of pandemic extinction, the model explains the spread at the global level observed in the 2009 H1N1 pandemic as induced by the interplay between the heterogeneity of the air mobility network structure, favoring the spread, and the population partition. A major epidemic is always achieved for $R_0 \geq 1.2$ even in the case children are assumed not to travel, when the model is parameterized with European countries data and statistics. Results are also consistent with the occurrence of sporadic outbreaks in continental Europe during summer 2009 and widespread transmission in the UK, once the model is informed with the substantial reduction in transmission associated to school holidays.

Despite the presence of various other epidemiological factors that may influence the epidemic outcome, our results suggest that the variations of demographic and mixing profiles across countries are an important source of heterogeneity in the epidemic outcome. This applies in particular to the contacts ratio that is observed to vary significantly and to have a large impact on the invasion potential. Such findings calls for the need to develop further studies in order to identify the social factors that affect this parameter and design targeted interventions, such as work-related measures, that may lower it, thus reducing the risk of an outbreak.

Given the availability of data regarding demographic, mixing and travel profiles, the model results can be used to assess the risks of a given outbreak scenario in a specific country for a newly emerging pathogen. Collecting data on population partitions and mixing matrices or developing alternative methods to estimate the contact patterns based on the demographic information available [25,74] is therefore important to make this approach applicable to a larger range of scenarios, as shown with the example of the US based on synthetic contact information [25].

Though based on simplifying assumption, the model is able to integrate the heterogeneities in the spatial distribution of the population, in the mixing patterns and in the travel behavior, and provide a solution to assess the risk of a major epidemic. We considered a definition for the children age class up to 15 or 18 years old, justified by the available data and statistics, however the approach is transparent to this choice and analogous results to the ones presented can be reached by informing the model with a different definition of classes, as long as statistics informing the groups-specific parameters $\alpha$, $\varepsilon$, $\eta$, and $r$ are available. The approach represents a general framework that can be applicable to other case studies and host population partitions that do not depend on age, such as for instance mixing patterns and travel behaviors depending on socio-economic aspects, or contact profiles and mobility within specific settings where classes correspond to professional roles or conditions of individuals (e.g. health-care workers and patients in hospitals).

**Acknowledgments**

The authors would like to thank Niel Hens for providing us the daily contact data of the POLYMOD study and for his useful comments; Caterina Rizzo for the age profile of the laboratory confirmed cases in Italy during summer 2009; Stephen Eubank for providing us the synthetic contact data for Portland; and Jose J Ramasco and Pablo Jensen for useful discussions. This work has been partially funded by the ERC Ideas contract no. ERC-2007-Stg204863 (EPIFOR) and the EC-Health contract no. 278433 (PREDEMICS) to VC and CP; the EU-FP7 contract no. 231807 (EPIWORK) to AA; the ANR contract no. ANR-12-MONU-0018 (HARMSFLU) to VC.

# References

1. Khan K et al.: **Spread of a Novel Influenza A (H1N1) Virus via Global Airline Transportation.** *New Engl J Med* 2009, **361**(2):212-4
2. Fraser C, Donnelly CA, Cauchemez S et al: **Pandemic potential of a strain of influenza A (H1N1): early findings.** *Science* 2009, **324**: 1557–1561.
3. Balcan D, Hu H, Goncalves B, Bajardi P, Poletto C, Ramasco JJ, Paolotti D, N. Perra, Tizzoni M, Van den Broeck W, Colizza V, Vespignani A: **Seasonal transmission potential and activity peaks of the new influenza A(H1N1): a Monte Carlo likelihood analysis based on human mobility.** *BMC Medicine* 2009, **7**:45.
4. Nishiura H, Castillo-Chavez C, Safan M, Chowell G: **Transmission potential of the new influenza A (H1N1) virus and its age-specificity in Japan**. *Euro Surveillance* 2009, **14**:19227.
5. **Novel swine-origin influenza A (H1N1) virus investigation team. Emergence of a novel swine-origin influenza A (H1N1) virus in humans.** *N Engl J Med* 2009, **360**:2605–2615
6. Cutler J, Schleihauf E, Hatchette TF, et al.: **Investigation of the first cases of human-to-human infection with the new swine-origin influenza A (H1N1) virus in Canada.** *CMAJ* 2009, **181**:159-63
7. Miller E, Hoschler K, Stanford E, Andrews N, et al: **Incidence of 2009 pandemic influenza A H1N1 infection in England: a cross-sectional serological study.** *Lancet* 2010, **375**: 1100-1108
8. Hahné S, Donker T, Meijer A, Timen A, van Steenbergen J, Osterhaus A, van der Sande M, Koopmans M, Wallinga J, Coutinho R: **The Dutch New Influenza A(H1N1)v Investigation Team. Epidemiology and control of influenza A(H1N1)v in the Netherlands: the first 115 cases.** *Eurosurveillance* 2009, **14**(27).
9. Belgian Working Group on influenza A(H1N1)v: **Influenza A(H1N1) virus infections in Belgium, May-June 2009.** *Eurosurveillance* 2009, **14**(28).

10. Health Protection Agency and Health Protection Scotland new influenza A(H1N1) investigation teams: **Epidemiology of the new influenza A(H1N1) in the United Kingdom,** April-May 2009
11. Influenza A(H1N1)v investigation teams: **Modified surveillance of influenza A(H1N1)v virus infections in France.** *Eurosurveillance* 2009, **14**(29).
12. Rizzo C, Declich S, Bella A, Caporali MG, Lana S, Pompa MG, Vellucci L, Salmaso S: **Enhanced epidemiological surveillance of influenza A(H1N1)v in Italy.** *Eurosurveillance* 2009, **14**(27).
13. Nishiura H: **Travel and age of influenza A (H1N1) 2009 virus infection,** *Journal of Travel Medicine* 2010, **17**(4): 269–270.
14. Nishiura H, Cook AR, Cowling BJ: **Assortativity and the probability of epidemic extinction: A case study of pandemic influenza A (H1N1-2009).** *Interdisciplinary Perspectives on Infectious Diseases* 2011, **2011**: 194507
15. Barrat A, Barthelemy M, Pastor-Satorras R, Vespignani A: **The architecture of complex weighted networks.** *Proc. Natl. Acad. Sci. USA* 2004, **101:** 3747–3752.
16. Guimera R, Mossa S, Turtschi A, Amaral LAN: **The worldwide air transportation network: Anomalous centrality, community structure, and cities global role.** *Proc. Natl. Acad. Sci. USA* 2005, **102:** 7794–7799.
17. DeMontis A, Barthelemy M, Chessa A, Vespignani A: **The structure of interurban traffic: a weighted network analysis.** E*nvironment Plann. B* 2007, **34**(5): 905 – 924.
18. Chowell G, Hyman JM, Eubank S, Castillo-Chavez C: **Scaling laws for the movement of people between locations in a large city.** *Phys. Rev. E* 2003, **68:** 066102**.**
19. Barrett CL, et al:**TRANSIMS: Transportation Analysis Simulation System.** Technical Report LA-UR-00-1725, Los Alamos National Laboratory 2000.
20. Eurostat, http://epp.eurostat.ec.europa.eu/portal/page/portal/eurostat/home/
21. **United Nations, Department of Economics and Social Affairs, Population Division, Population Estimates and Projections Sections,** [http://esa.un.org/unpd/wpp/Excel-Data/population.htm]
22. Mossong J, Hens N, Jit M, Beutels P, Auranen K, et al: **Social Contacts and Mixing Patterns Relevant to the Spread of Infectious Diseases.** *PLoS Med* 2008, **5**(3): e74.
23. Anderson RM, May RM: *Infectious Diseases of Humans: Dynamics and Control* Oxford:Oxford University Press; 1992.
24. Wallinga J, Teunis P, Kretzschmar M: **Using data on social contacts to estimate age-specific transmission parameters for respiratory-spread infectious agents.** *American Journal of Epidemiology* 2006, **164**: 936–944
25. Del Valle SY, Hyman J, Hethcote HW, Eubank SG: **Mixing patterns between age groups in social networks**. *Social Networks* 2007, **29**: 539–554.
26. Kiss I, Simon PL, Kao R**: A contact network-based formulation of a preferential mixing model.** *Bulletin of Mathematical Biology* 2009, **71** (4), 888–905
27. Goeyvaerts N, Hens N, Aerts M, Beutels P: **Model structure analysis to estimate basic immunological processes and maternal risk for parvovirus B19.** *Biostatistics* 2011, **12**(2):283-302.
28. Colizza V, Vespignani A: **Epidemic modelling in metapopulation systems with heterogeneous coupling pattern: Theory and simulations.** *J. Theor. Biol.* 2008, **251**: 450–467.
29. Pastor-Satorras R, Vespignani A: **Epidemic spreading in scale-free networks.** *Phys. Rev. Lett.* 2001, **86**: 3200–3203.

30. Colizza V, Pastor-Satorras R, Vespignani A: **Reaction-diffusion processes and metapopulation models in heterogeneous networks.** *Nature Phys*. 2007,**3:** 276–282
31. Goeyvaerts N, Hens N, Ogunjimi B, Aerts M, Shkedy Z, Van Damme P, Beutels P: **Estimating infectious disease parameters from data on social contacts and serological status.** *Journal of the Royal Statistical Society* 2010, **59**:255
32. Goldstein E, Apolloni A, Lewis B, Miller JC, Macauley M, Eubank S,Lipsitch MJ, Wallinga J: **Distribution of vaccine/antivirals and the "least spread line" in a stratified population.** Journal of the Royal Society Interface 2010, **7**: 755–764.
33. Diekmann O, Heesterbeek JAP, Roberts MG: **The construction of next-generation matrices for compartmental epidemic models**. *Journal of the Royal Society Interface* 2010, **7**: 873–885.
34. Leung GM, Hedley AJ, Ho LM, Chau P, Wong IO, Thach TQ et al (2004) **The epidemiology of severe acute respiratory syndrome in the 2003 Hong Kong epidemic: an analysis of all 1755 patients**. *Ann Intern Med* **141**, 662-673.
35. Nishiura H, Chowell G, Safan M, Castillo-Chavez C: **Pros and cons of estimating the reproduction number from early epidemic growth rate of influenza A (H1N1) 2009**. *Theoretical Biology and Medical Modelling* 2010, **7**(1):1.
36. Eames KTD, Tilston NL, Brooks-Pollock E, Edmunds WJ: **Measured Dynamic Social Contact Patterns Explain the Spread of H1N1v Influenza.** *PLoS Comput Biol* 2012, **8**(3): e1002425.
37. Hens N, Ayele G M, Goeyvaerts N, Aerts M, Mossong J, Edmunds JW, Beutels P: **Estimating the impact of school closure on social mixing behaviour and the transmission of close contact infections in eight European countries.** *BMC Infectious Diseases* 2009, **9**:187
38. Diekmann O, Heesterbeek JAP, Metz JAJ: **On the definition and the computation of the basic reproduction ratio r0 in models for infectious diseases in heterogeneous populations.** *Journal of Mathematical Biology* 1990, **28**: 365–382 (1990).
39. Bajardi P, Poletto C, Ramasco JJ, Tizzoni M, Colizza V, Vespignani A: **Human Mobility Networks, Travel Restrictions, and the Global Spread of 2009 H1N1 Pandemic.** *PLoS ONE* 2011, **6**(1): e16591.
40. Ikonen N, Strengell M, Kinnunen L, Osterlund P, Pirhonen J, Broman M, Davidkin I, Ziegler T, Julkunen I: **High frequency of cross-reacting antibodies against 2009 pandemic influenza A(H1N1) virus among the elderly in Finland.** *Euro Surveillance*. 2010,**15**:19478.
41. Hancock K, Veguilla V, Lu X, Zhong W, Butler EN, Sun H, Liu F, Dong L, DeVos JR, Gargiullo PM, Brammer TL, Cox NJ, Tumpey TM, Katz JM: **Cross-reactive antibody responses to the 2009 pandemic H1N1 influenza virus**. *N. Engl. J. Med.* 2009, **361**:1945-1952.
42. [Allwinn-2010] Allwinn R, Geiler J, Berger A, Cinatl J, Doer HW**: Determination of serum antibodies against swine-origin influenza A virus H1N1/09 by immunofluorescence, haemagglutination inhibition, and by neutralization tests: how is the prevalence rate of protecting antibodies in humans?** *Medical Microbiology Immunology* 2010, **199**:117-121.
43. Ball F, Mollison D, Scalia-Tomba G: **Epidemics with two levels of mixing.** *Ann. Appl. Probab.* 1997, **7**: 46–89.
44. Cross P, Lloyd-Smith JO, Johnson PLF, Wayne MG: **Duelling timescales of host movement and disease recovery determine invasion of disease in structured populations.** *Ecol. Lett.* 2005, **8**: 587–595.

45. Cross P, Johnson PLF, Lloyd-Smith JO, Wayne MG **Utility of R0 as a predictor of disease invasion in structured populations.** *J. R. Soc. Interface* 2007, 4: 315–324.
46. Colizza V, Vespignani A: **Invasion threshold in heterogeneous metapopulation networks.** *Phys. Rev. Lett.* 2007, **99**: 148701.
47. Balcan D, Vespignani A: **Phase transitions in contagion processes mediated by recurrent mobility patterns.** *Nature Phys.* 2011, **7**: 581–586.
48. Belik V, Geisel T, Brockmann D: **Natural human mobility patterns and spatial spread of infectious diseases.** *Phys. Rev. X* 2011, **1**:011001.
49. Meloni S, Perra N, Arenas A, Gomes S, Moreno Y, Vespignani A: **Modeling human mobility responses to the large-scale spreading of infectious diseases.** *Scientific Reports* 2011, **1**:62.
50. Poletto C, Tizzoni M, Colizza V: **Heterogeneous length of stay of hosts' movements and spatial epidemic spread.** *Nature Scientific Reports* 2012, **2**: 476.
51. Harris TE: *The Theory of Branching Processes.* Mineola: Dover Publications;1989.
52. Ball F, Clancy D: **The final size and severity of a generalised stochastic multitype epidemic model.** *Advances in Applied Probability* 1993, **25**(4):721–736.
53. Flasche S, Hens N, Boëlle PY, Mossong J, van Ballegooijen WM, Nunes B, Rizzo C, Popovici F, Santa-Olalla P, Hrubá F, Parmakova K, Baguelin M, van Hoek AJ, Desenclos JC, Bernillon P, Cámara AL, Wallinga J, Asikainen T, White PJ, Edmunds WJ: **Different transmission patterns in the early stages of the influenza A(H1N1)v pandemic: a comparative analysis of 12 European countries.** *Epidemics* 2011, **3**(2):125-33.
54. Mills CE, Robins JM, Lipsitch M: Transmissibility of the 1918 pandemic influenza. *Nature* 2004, 432:904-906.
55. Chao DL, Halloran ME, Obenchain VJ, Longini IM: **FluTE a publicly available stochastic influenza epidemic simulation model.** *PLoS Comput Biol* 2010, **6**:e1000656.
56. Watts DJ, Muhamad R, Medina DC, Dodds PS: **Multiscale, resurgent epidemics in a hierarchical metapopulation model.** *Proc Natl Acad Sci USA* 2005, **102:**11157-11162
57. Epstein JM, Goedecke DM, Yu F, Morris RJ, Wagener DK, Bobashev GV: **Controlling Pandemic Flu: The Value of International Air Travel Restrictions.** *PLoS ONE* 2007, **2**: 1-11.
58. Cooper BS, Pitman RJ, Edmunds WJ, Gay NJ: **Delaying the international spread of pandemic influenza.** PLoS Medicine 2006, **3**: 845-855.
59. Hollingsworth TD, Ferguson NM, Anderson RM: **Will travel restrictions control the international spread of pandemic influenza?** *Nature Medicine* 2006, **12**: 497–499.
60. Lam EHY, Cowling BJ, Cook AR, Wong JYT, Lau MSY, Nishiura H: **The feasibility of age-specific travel restrictions during influenza pandemics.** Theoretical Biology and Medical Modelling 2011, **8**: 44-57.
61. Merler S, Ajelli M: **The role of population heterogeneity and human mobility in the spread of pandemic influenza.** *Proceedings of the Royal Society B: Biological Sciences* 2010, **277**: 557–565.
62. ECDC Technical report **Guide to public health measures to reduce the impact of influenza pandemics in Europe: 'The ECDC Menu'** (2009).

63. Cowling BJ, Lau LLH, Wu P, Wong HWC, Fang VJ, Riley S, Nishiura H: **Entry screening to delay local transmission of 2009 pandemic influenza A(H1N1).** *BMC Infect Dis 2010*, **10**:82.
64. Eubank S, Guclu H, Anil Kumar VS, Marathe MV, Srinivasan A, Toroczkai Z, Wang N (2004) **Modelling disease outbreaks in realistic urban social networks.** *Nature* **429:** 180-184.
65. Halloran ME, Ferguson NM, Eubank S, Longini IM, Cummings DAT, Lewis B, Xu S, Fraser C, Vullikanti A, Germann TC, Wagener D, Beckman R, Kadau K, Macken , Burke DS, Cooley P (2008) **Modeling targeted layered containment of an influenza pandemic in the United States.** *Proc Natl Acad Sci USA* **105:** 4639-4644.
66. Keeling MJ, Grenfell BT: **Effect of variability in infectious period on the persistence and spatial spread of infectious diseases.** *Math Biosci* 1998,**147**: 207–226.
67. Vergu E, Busson H, Ezanno P: **Impact of the Infection Period Distribution on the Epidemic Spread in a Metapopulation Model.** *PLoS ONE* 2010, **5**(2): e9371.
68. Tizzoni M, Bajardi P, Poletto C, Ramasco JJ, Balcan D, Goncalves B, Perra N, Colizza V, Vespignani A: **Real-time numerical forecast of global epidemic spreading: case study of 2009 A/H1N1pdm**. *BMC Medicine* 2012, 10:165.
69. Funk S, Salathe M, Jansen VAA: **Modeling the influence of human behavior on the spread of infectious diseases: a review.** *J Roy Soc Interface* 2010.
70. Salathé M, Khandelwal S: **Assessing Vaccination Sentiments with Online Social Media: Implications for Infectious Disease Dynamics and Control.** *PLoS Comput Biol* 2011, **7**(10): e1002199.
71. Perra N, Balcan D, Goncalves B, Vespignani A: **Towards a characterization of behavior-disease models.** *PLOS One* 2011, **6**(8):e23084
72. Bagnoli F, Lio P, Sguanci L: **Risk perception in epidemic modeling**. *Phys Rev E* 2007, 76: 061904.
73. Poletti P, Caprile B, Ajelli M, Pugliese A, Merler S: **Spontaneous behavioural changes in response to epidemics.** *J Theor Biol* 2009, 260:31.
74. Fumanelli L, Ajelli M, Manfredi P, Vespignani A, Merler S: **Inferring the Structure of Social Contacts from Demographic Data in the Analysis of Infectious Diseases Spread.** *PLoS Comput Biol* 2012, 8(9): e1002673.

# Tables

**Table 1:** Variables used to define the age classes in the epidemic model; $c =$children, $a =$adults.

| variable | definition |
| --- | --- |
| $\alpha$ | children fraction of the population |
| $q_a, q_c$ | average number of contacts per unit time established by individuals in the children and adult classes, respectively |
| $\eta = q_a/q_c$ | ratio of the average number of contacts |
| $\varepsilon = \varepsilon_c \alpha = \eta \varepsilon_a (1 - \alpha)$ | total fraction of contacts across age classes |
| $r$ | children fraction of the traveling population |

**Table 2:** Values of the age classes parameters obtained from country-specific statistics and data (see main text for references).

| country | $\alpha$ | $\eta$ | $\varepsilon$ |
| --- | --- | --- | --- |
| Belgium | 0.21 | 1.13 | 0.125 |
| Germany | 0.18 | 0.75 | 0.098 |
| Finland | 0.21 | 0.79 | 0.091 |
| Great Britain | 0.22 | 0.75 | 0.115 |
| Italy | 0.17 | 0.62 | 0.083 |
| Luxembourg | 0.22 | 0.93 | 0.107 |
| The Netherlands | 0.22 | 0.83 | 0.094 |
| Poland | 0.21 | 0.97 | 0.100 |
| *Europe (average values)* | *0.20* | *0.79* | *0.097* |
| Mexico | 0.32 | 0.32 | 0.063 |

**Table 3:** Variables defining the metapopulation model with one class of individuals in the degree-block approximation.

| variable | definition |
| --- | --- |
| $k$ | degree of a subpopulation, i.e. number of airline connections to other subpopulations |
| $V, V_k$ | total number of subpopulations, number of subpopulations with degree $k$ |
| $N_k$ | population size of subpopulations with degree $k$ |
| $w_{kk'} = w_0 (kk')^\theta$ | number of passengers flying from a subpopulation with degree $k$ to a subpopulation with degree $k'$ |
| $w_0$ | mobility scale |
| $d_{kk'} = w_{kk'}/N_k$ | diffusion rate of passengers flying from a subpopulation with degree $k$ to a subpopulation with degree $k'$ |

# Figures legends

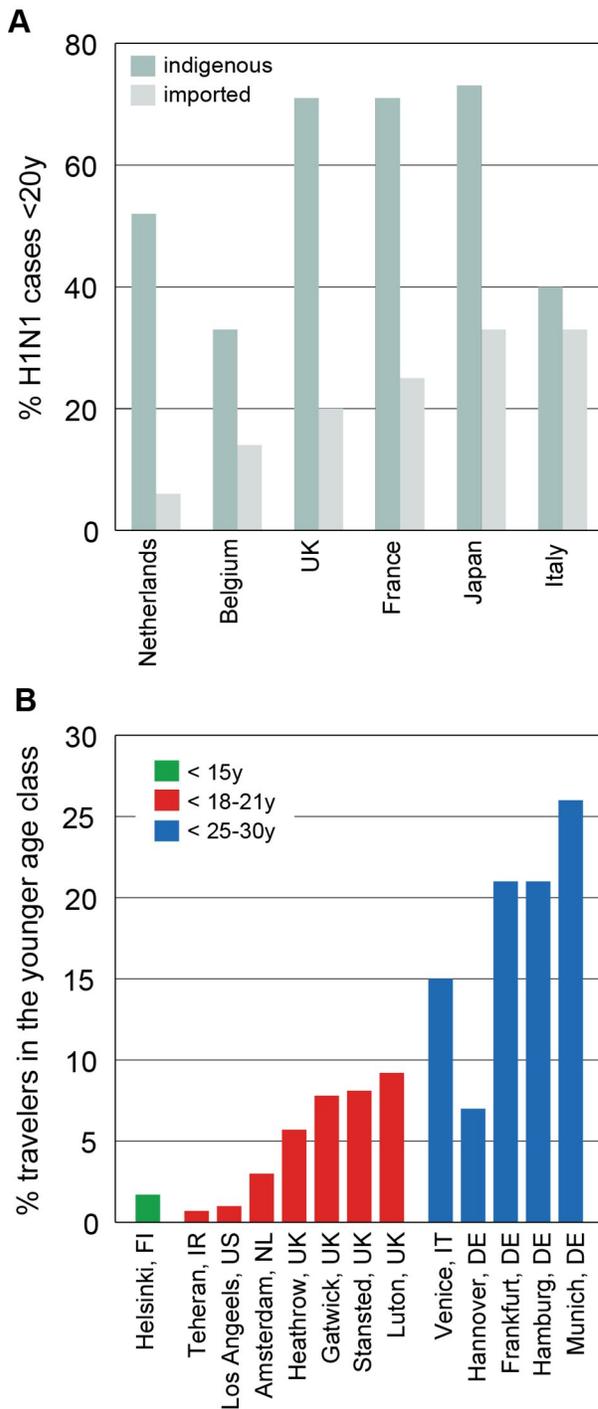

**Figure 1: Imported vs. indigenous H1N1 cases and age-specific travel statistics for various countries.** (a) Fraction of indigenous cases and of imported cases during the initial phase of the H1N1 pandemic outbreak in the [0-19] years age class, calculated from surveillance data for the following countries: The Netherlands [8], Belgium [9], UK

[10], France [11], Japan [13], Italy [12]. (b) Percentage of air-travel passengers in the younger age classes for a set of airports around the world. The age classification used by the demographic statistics vary across countries (Helsinki[1], Finland; Teheran[3], Iran; Los Angeles[2], USA; Amsterdam[5], The Netherlands; Heathrow[4], Gatwick[4], Stansted[4], Luton[4], UK; Venice[6], Italy; Hannover[7], Frankfurt[8], Hamburg[8], Munich[8], Germany) with the corresponding age brackets for the children class (expressed in years): 1=[0,15]; 2=[0,18]; 3=[0,19]; 4=[0,20]; 5=[0,21]; 6=[0,25]; 7=[0,26]; 8=[0,30]. Sources of the data are reported in the Additional File. The statistics found for Italy and Germany correspond to larger age brackets. If we rescale the data as indicated in the main text, we obtain the following estimates for the percentage of travelers in the [0-18] years old class: 1.05% (Venice), 0.49% (Hannover), 2.31% (Frankfurt), 2.31% (Hamburg), and 2.86% (Munich). These estimated values are consistent with the data.

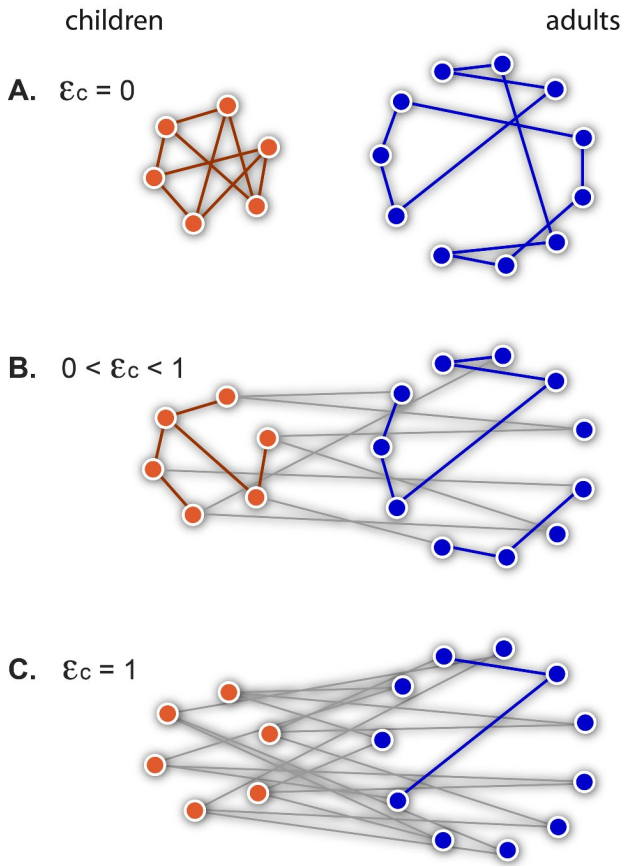

**Figure 2: Schematic example of different assortativity levels in mixing patterns.** Throughout the paper we use $\varepsilon = \varepsilon_c \alpha$ as the parameter referring to the assortativity of the mixing pattern, since it represents the total fraction of across-groups contacts,. In this scheme we show three examples of different assortativity levels. A: maximum assortativity, corresponding to no mixing between the two classes ($\varepsilon = \varepsilon_c = 0$); B: intermediate assortativity, i.e. a given fraction of the children contacts are directed to adults (like e.g. a random mixing scenario), the others being of the child-child type; C: no assortativity in the children age class, as all contacts established by children are directed to the adults class ($\varepsilon_c = 1$ and thus $\varepsilon = \alpha$).

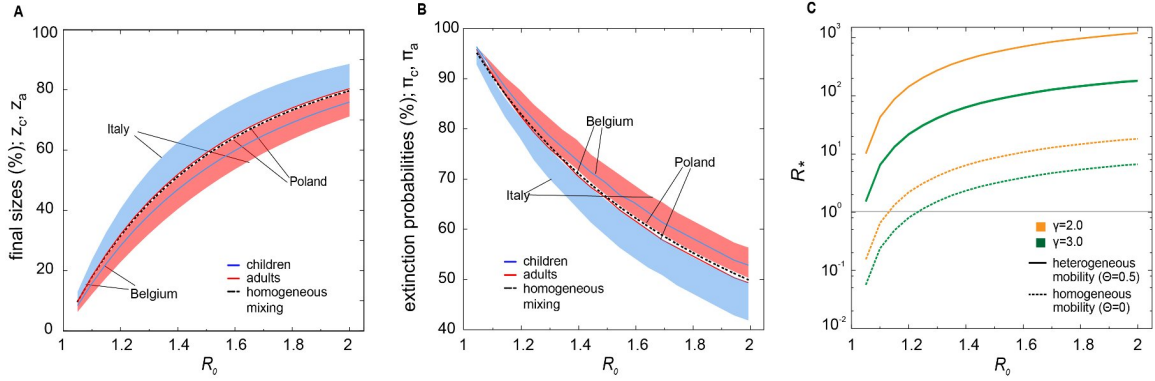

**Figure 3: Final size, extinction probability, and global invasion threshold vs. $R_0$.** A-B: Final sizes and extinction probabilities per age class as functions of the reproductive number $R_0$. The various curves for the eight European countries under study are shown by means of a shaded area, for the sake of visualization, with the exception of Belgium, see below. The maximum value for the epidemic size in children (and minimum for the epidemic size in adults) is obtained for Italy; the opposite (i.e. minimum $z_c$ and maximum $z_a$) is obtained for Poland. The situation is reversed for the extinction probabilities – the maximum value for the extinction probability in children (and minimum for the extinction probability in adults) is obtained for Poland; the opposite (i.e. minimum $\pi_c$ and maximum $\pi_a$) is obtained for Italy. In both plots, Belgium is a standalone example, with $z_c > z_a$ and $\pi_c > \pi_a$, differently from all other countries and due to the fact that it is the only population in the dataset to have $\eta > 1$, as discussed in the main text. The dashed line represents the case of homogeneous mixing when no partition of the population is considered (in panel B it corresponds to the function $R_0^{-1}$). C: Global invasion threshold $R_*$ as a function of the reproductive number $R_0$, for different values of the parameters describing the mobility process. Air mobility networks having degree distributions $P(k) \propto k^{-\gamma}$ with $\gamma = 2$ and $\gamma = 3$ are shown to consider different levels of heterogeneity. The results obtained in the two cases are compared to the scenarios with homogeneous diffusion rates $d_{kk'} = w_0(kk')^\theta$ obtained for $\theta = 0$. All curves are obtained by setting the fraction of passengers in the children class equal to the observed data, i.e. $r = 7\%$, and informing the model with the European average values for $\alpha, \eta, \varepsilon$.

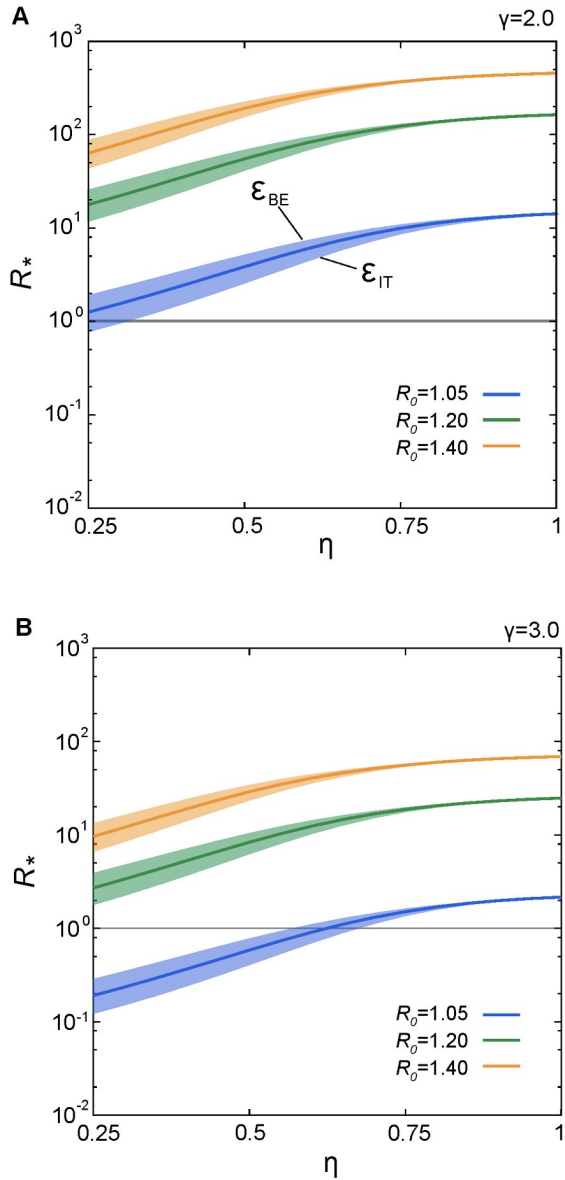

**Figure 4: Impact of contacts ratio $\eta$.** Global invasion threshold $R_*$ as a function of the contacts ratio $\eta$, for the case of a mobility air network structure having $P(k) \propto k^{-\gamma}$ with $\gamma = 2$ (panel A) and $\gamma = 3$ (panel B). Each plot considers three values of the reproductive number – $R_0 = 1.05, 1.2, 1.4$. The various curves for the eight European countries under study are shown by means of a shaded area, for the sake of visualization. The distribution of the population into the children class is set to the average European value for all countries, whereas the assortativity level is left country-specific. Maximum assortativity (and thus minimum $\varepsilon$) is reached for Italy, the opposite observed for Belgium. Here we assume that $r = 0$. The dashed line indicates the threshold condition $R_* = 1$.

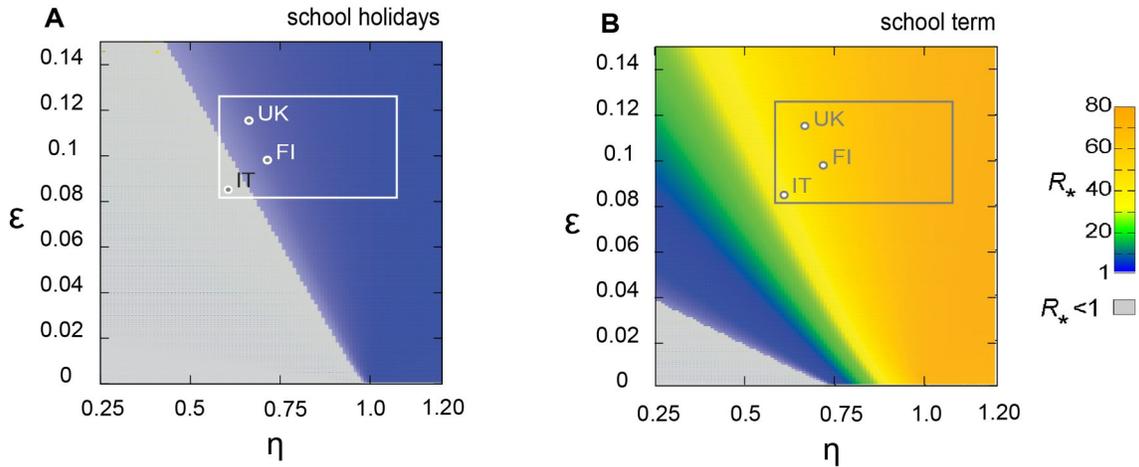

**Figure 5: Impact of assortativity and of school term vs. school holidays.** Global invasion threshold $R_*$ as a function of the across-groups mixing $\varepsilon$ and of contacts ratio $\eta$, for the reproductive number estimated during school holidays ($R_0 = 1.05$, panel A) and school term ($R_0 = 1.4$, panel B), based on contact data in the UK [62]. Here we fix the children fraction of the population $\alpha$ to its European average value. The grey area indicates the extinction phase where $R_* < 1$, whereas the colored area refers to the region of the parameter phase space that is above the threshold condition. The rectangular box indicates the area corresponding to the European intervals for the parameters $\varepsilon$ and $\eta$. The cases for Italy ($\varepsilon = 0.08, \eta = 0.62$), United Kingdom ($\varepsilon = 0.11, \eta = 0.75$), and Finland ($\varepsilon = 0.09, \eta = 0.79$) are highlighted.

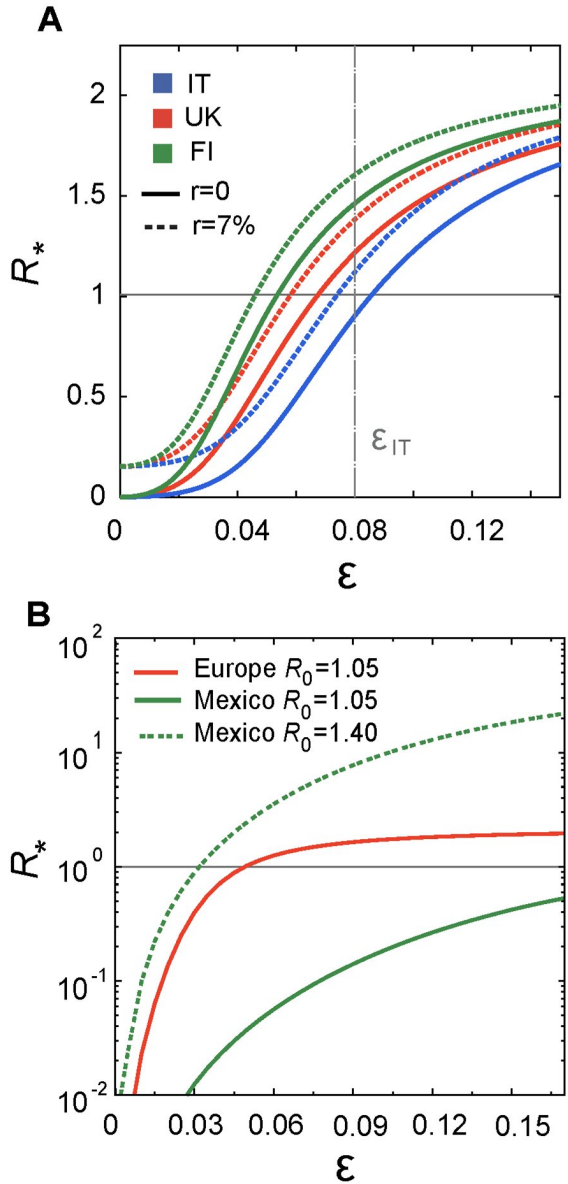

**Figure 6: Impact of age-specific travel behavior and age profile.** A: Global invasion threshold $R_*$ as a function of the across-groups mixing $\varepsilon$ for the cases of Italy, United Kingdom, and Finland, assuming $R_0 = 1.05$. The solid colored lines correspond to the cases when only adults travel ($r = 0$) and the dashed colored lines to the cases when a percentage of 7% of passengers belongs to the children class. The continuous horizontal line indicates the threshold condition $R_* = 1$. B: Global invasion threshold $R_*$ as a function of the across-groups mixing $\varepsilon$: comparison between Europe ($\alpha = 0.20, \eta = 0.79$) and Mexico ($\alpha = 0.32, \eta = 0.32$). Here we consider $R_0 = 1.05$ and $R_0 = 1.4$, assuming $r = 0$.

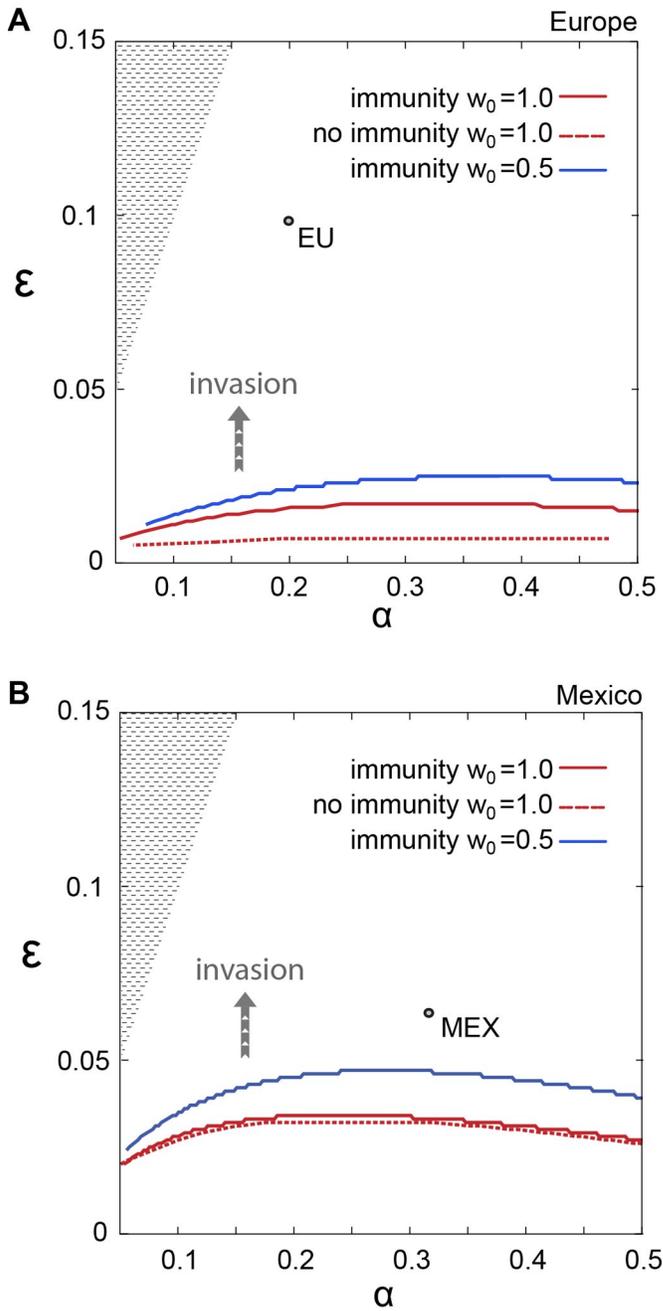

**Figure 7: Case with immunity.** Threshold condition $R_* = 1$ as a function of the across-groups mixing $\varepsilon$ and of the children fraction $\alpha$ for Europe (panel A) and Mexico (panel B): comparison of the no-immunity case with the case of pre-existing immunity and of travel reduction, modeled by setting $w_0 = 0.5$, consistently with the empirically observed drop to/from Mexico during the early stage of the 2009 H1N1 pandemic [37]. Here we consider: $R_0 = 1.2$ in Europe and $R_0 = 1.4$ in Mexico, i.e. the lower bound of the reproductive number estimated for the country from the initial outbreak data [2]. All travelers are adults ($r = 0$). The three lines, continuous red, dashed red and continuous blue, correspond to pre-existing immunity, no-immunity and travel reduction, respectively. Global epidemic invasion region is above each critical curve. The patterned

gray area refers to the region of parameter values that do not satisfy the consistency relation $\varepsilon < \min\{\alpha, \eta(1-\alpha)\}$.